# On the structure of systematic perturbation theory with unstable fields[1]

**Contribution to the XIII Int. Workshop on Quantum Field Theory and High Energy Physics
(QFTHEP'99, 27 May – 2 June 1999, Moscow)**

F. V. Tkachov [a]

*Institute for Nuclear Research of Russian Academy of Sciences
60th October Ave. 7a, Moscow 117312, Russia*

[a] ftkachov@ms2.inr.ac.ru

*Abstract.* Discussed is the structure of non-trivial counterterms that occur in the systematic gauge-invariant perturbation theory with unstable fields introduced in [hep-ph/9802307].

I would like to discuss the structure of the systematic perturbation theory for models with unstable fundamental fields (sysPT) proposed in [1]. The discussion below is intended to complement [1] to which I will freely refer.

Some general warnings are in order.

The industry of PT calculations for LEP etc. has been dealing with unstable particles in a more or less successful fashion for some time [2] (see also refs. in [1]) — but it has also been encountering difficulties (most notably, a lack of gauge invariance; see e.g. [3]). It is clear that to avoid groping in the dark one should have a clear formulaic understanding of the nature of the difficulties being encountered.

> The purpose of sysPT is to clarify in a systematic manner the mathematical nature of the weak-coupling limit in situations with unstable fields.

Just how, exactly, such an understanding may lead to improved calculations for LEP2 etc., I don't care at this point; I'm just happy to have obtained the understanding.[1]

Some physicists may be uncomfortable with the adjective "mathematical" above. But we are dealing here with the structures encountered in higher-order corrections and relevant for precision calculations, and the second and higher decimal digits in theoretical predictions cannot be accounted for by simple visualizations. So the kind of intuition needed to understand the sysPT concerns the structure of the formalism and therefore requires a (non-standard) mathematical expression.

Also recall the industry of multiloop calculations that emerged thanks to the clarification of the short-distance and mass expansions[2] within the framework of the method of asymptotic operation[3]. The calculational efficiency is due to the fact that the method of AO yields expansions in a maximally simple form, i.e. the resulting expansions run in pure powers and logarithms of the expansion parameter[4]. This allows to develop highly efficient calculational algorithms geared towards automated large-scale calculations[5].

The sysPT implements the same logic: it is based on the same powerful method of AO[6], and it yields the expansions (now the expansion parameter is the coupling) in a maximally simple form. The form is somewhat unconventional (VP-distributions are involved) but is otherwise rather simple. This, one may hope, provides a starting point for developing more efficient calculational procedures than the currently employed.

Now on to the sysPT.

The key difficulty in the construction of PT with unstable fields is that the formal expansion of amplitudes in powers of the coupling yields singular expressions which, when squared to form probabilities, result in expressions that are non-integrable at the zero-coupling mass shell of unstable particles. But integrability is a key physical requirement (taking into account, say, initial state radiation amounts to a convolution with suitably chosen kernels [2]).

All conventional approaches to resolution of this difficulty abandon the idea of a complete expansion in powers of the coupling: self-energies are left in the denominators, and then one performs a more or less sophisticated massaging of the resulting expressions aimed at obtaining manageable gauge-invariant theoretical predictions.

The drawback of such an approach is that it involves huge amounts of hand work, and there is an opinion rather unambiguously expressed by some experts involved in the LEP1 calculations [2] that it is impossible with the old techniques to accomplish anything similar to the complete one-loop level LEP1 calculations for LEP2 (at LEP2 one deals with $O(10^4)$ one-loop diagrams [9]). It is clear that what one needs

---

[1] Some related materials can be found at
http://www.inr.ac.ru/~ftkachov/projects/unstable/index.htm.

[2] For a review see [4], [5].

[3] For a review and references to the Euclidean variant of the theory of AO see [4].

[4] The so-called property of perfectness [6]; also note that such expansions are essentially unique [4].

[5] With short-distance and mass expansions, it is the celebrated integration-by-parts algorithms [7].

[6] The required extension of the Euclidean AO to arbitrary asymptotic regimes in Minkowski space was achieved in [8], thus yielding a much sought solution of the general asymptotic expansion problem in perturbative QFT. The recipes of [8] are valid for both loop and phase-space integrals because the distribution-theoretic nature of AO makes it insensitive to whether the individual factors are ordinary propagators or phase-space $\delta$-functions.

here is a very systematic, "mechanical" approach that would allow a high level of automation.

The sysPT offers a new route to this goal. One observes that the difficulties of the conventional approaches reduce, from mathematical viewpoint, to a non-commutativity of the perturbative expansion in powers of the coupling and the operation of squaring of amplitudes to obtain observed probabilities. Once this is realized, it is natural to attempt to obtain an expansion directly for probabilities.

First one recalls the elementary formula ($x$ is a real argument and $\Gamma$ is small):

$$\frac{1}{x^2 + \Gamma^2} \underset{\Gamma \to 0}{=} \frac{1}{\pi \Gamma} \delta(x) + O(1) . \qquad 0.1$$

This is easily verified e.g. by explicit integrations in infinite limits with rational functions that do not have zeros near $x = 0$ such as $(c + ix)^{-n}$.

From Eq. 0.1 one obtains

$$\left| \frac{1}{M^2 - Q^2 - g^2 \Sigma(Q^2)} \right|^2$$

$$\underset{g \to 0}{=} g^{-2} \pi \delta(M^2 - Q^2) \frac{1}{\operatorname{Im}\Sigma(M^2)} + O(1) . \qquad 0.2$$

Here $M$ is the Lagrangian mass of the unstable particle (call it $X$), $\Sigma$ is its one-loop self-energy (Dyson-resummed into denominators), and $g$ is the coupling responsible for the instability.

Eq. 0.2 is nothing but a well-known relation in disguise:

$$\sigma(q_1 \bar{q}_2 \to X \to l_1 \bar{l}_2) \approx \sigma(q_1 \bar{q}_2 \to X) \times Br(X \to l_1 \bar{l}_2) . \qquad 0.3$$

Indeed, the $\delta$-function on the r.h.s. of 0.2 describes free propagation of $X$ in the final state. So Eqs. 0.2 and 0.3 only describe the fact that the unstable particle becomes stable in the zero-coupling limit. This is a fundamental boundary condition for any systematic expansion of probabilities in powers of the coupling. We see:

(i) The exact expansion is bound to contain anomalous $\delta$-functional terms. The simplest such term (the one shown on the r.h.s. of 0.2) has a fundamental physical meaning.

(ii) The $\delta$-functional terms spoil the naïve counting of powers of the coupling. Indeed, the l.h.s. of 0.2 corresponds to tree + one-loop approximation and one would normally expect the r.h.s. to contain $O(1)$ and $O(g^2)$ terms. However, the coefficient of $\delta$-function contains an $O(g^{-2})$ contribution from the one-loop of self-energy. Because only the imaginary part are involved, this is equivalent to the total width of the decay of $X$ in the lowest order (the decay vertex is taken in the tree approximation, and the cut loop comes from phase space).

(iii) It is clear that if one includes two-loop $O(g^4)$ terms into the self-energy ($g^2 \Sigma \to g^2 \Sigma_{1-\text{loop}} + g^4 \Sigma_{2-\text{loop}}$) then it is safe to Taylor-expand in all occurrences of $g$ except the one corresponding to the one-loop self-energy. The two-loop contributions to self-energy contribute at the level $O(1)$ in the r.h.s. Since only imaginary parts will contribute, this corresponds to one-loop total width (one-loop decay vertices and another loop from phase space of decay products). This anomaly of power counting for the coupling has long been felt to occur (cf. the discussion in [3]) but a systematic method of book-keeping has been lacking.

(iv) Finally, there is nothing mysterious about the anomalous $\delta$-functional terms: the simplest one has a clear physical meaning, and the higher ones are simply corrections that form a more or less regular pattern.

For the squared propagator (the l.h.s. of 0.2) the expansion problem reduces to a one-dimensional expansion problem which generalizes 0.1, and many felt that corrections to Eq. 0.2 are bound to contain derivatives of the $\delta$-function. The method of AO (in particular, the secondary so-called "homogenization" expansion [8]) make obtaining such expansions a mechanical procedure.

The expansion 0.2 can be pushed to all orders in $g$: see [1] for explicit expressions. For the purposes of illustration I show here only the $O(1)$ term:

$$\operatorname{VP}\left[(M^2 - Q^2)^{-2}\right] + Z_1 \delta(M^2 - Q^2) - Z_2 \delta'(M^2 - Q^2), \qquad 0.4$$

where $Z_i$ can be explicitly expressed in terms of the coefficients of the Taylor expansion of $\Sigma(Q^2)$ at $Q^2 = M^2$.

Eq. 0.4 is to be compared with the naïve expression obtained by squaring the tree-level unstable particle's propagator:

$$(M^2 - Q^2)^{-2} . \qquad 0.5$$

This has a non-integrable singularity at $Q^2 = M^2$.

We see that the correct expression (Eq. 0.4) differs from the naïve one (Eq. 0.5) in two respects:

(i) In the correct expansion, the non-integrable singularity of 0.5 regulated by the VP prescription (see [1] for a definition). This only affects the naïve expression 0.5 exactly at $Q^2 = M^2$. This may seem unusual but such things are common in the theory of distributions: an integrable distribution can be obtained from a non-integrable singular function by modifying the latter only exactly at the point of singularity.

(ii) On top of the VP prescription, there are $\delta$-functional counterterms. The presence of the first derivative of the $\delta$-function correlates with the fact that the singularity of 0.5 is linear by power counting.

The described pattern (an intermediate regularization plus a fine-tuning by $\delta$-functional "counterterms") is very general and occurs systematically in the theory of AO. The first time this pattern occurred in the theory of UV renormalization by Bogolyubov [10], and it is instructive to compare the structure of 0.4 with the $R$-operation.

The logical structure of the reasoning that led to sysPT essentially follows Bogolyubov's theory of UV renormalization [10]. In both cases:

(i) The difficulty is traced back to an incorrect formal manipulation which ignores the generalized nature of the mathematical objects involved. In the case of UV divergences, it is the formal multiplication of singular functions (propagators in chronological products). In the case of unstable particles, it is an unjustified expansion of amplitudes into a series in powers of the coupling prior to squaring.

(ii) The final result is required to be a locally integrable distribution (in the space of coordinates in the case of $R$-operation, and in the aggregate space of loop and phase space momenta in the case of sysPT) and it is observed that to fix its structure at the point of singularity, one needs a special procedure.

(iii) The final answer has a characteristic form "naïve expression + regularization + counterterms". The counterterms are proportional to $\delta$-functions (and their derivatives) localized exactly at the points of singularity, and the number of derivatives of $\delta$-functions is determined by a power counting.

There are also differences:

- UV singularities are localized on flat manifolds (linear subspaces) in the space of coordinates whereas the singularities due to unstable particles are localized on non-linear manifolds (the mass shell $Q^2 = M^2$ in the simplest case of 0.5).

- The finite parts of the UV counterterms of the $R$-operation are arbitrary whereas the coefficients of the $\delta$-functional counterterms of sysPT are uniquely fixed: they are explicitly expressed in terms of special integrals obtained via the so-called consistency conditions of AO supplemented by the homogenization procedure (a special secondary expansion designed to yield an expansion of a purely power-and-log type).

Now the expansion of the squared propagator 0.2 is far from the whole story. The point is, the singularities due to instabilities (i.e. the singularities of the expressions 0.5 and 0.4) may interact with the singularities of other factors in the diagram such as photon, gluon, etc. propagators. This leads to more complex singular configurations that also require addition of the corresponding counterterms. The method of AO [4], [8] offers a systematic way to determine such counterterms.

In the remaining part of this talk I would like to discuss the structure of such counterterms.

The simplest example (a configuration with one photon line attached by both ends to the unstable propagator) was discussed in [1]. The observed features: the coefficient of the corresponding $\delta$-functional counterterm contains a logarithm of the coupling; such logarithms cancel out in the sum over the corresponding gauge-invariant subset of diagrams; this cancellation was traced to a mechanism completely similar to that behind the well-known cancellations of IR singularities in QED.

In more complex cases the singular configurations of propagators that require introduction of non-trivial counter-terms are grouped into families, with the diagrams within a family differing by the points on the chains of (unstable) propagators to which massless propagators are attached (cf. gauge-invariant subsets of diagrams in QED). An example of such a family is as follows:

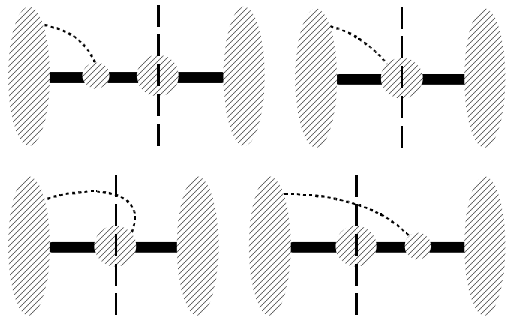

0.6

The fat lines correspond to unstable propagators (forming chains in the sense of [1]), there is a "photon" line, and the blobs correspond to subgraphs which do not belong to the singular configuration. Note that the singular subgraphs need not be connected subgraphs in the usual graph-theoretic sense.

At this point to avoid confusion I would like to emphasize that the diagrammatic images that emerge here are entirely determined by the underlying analytical structures, so there need not be any direct correspondence with, say, singular subgraphs and any standard graph-theoretic notions invented a priori.

The following figure represents some more families of singular configurations; each diagram depicts only one member of a family, with the total number of the diagrams shown as a factor:

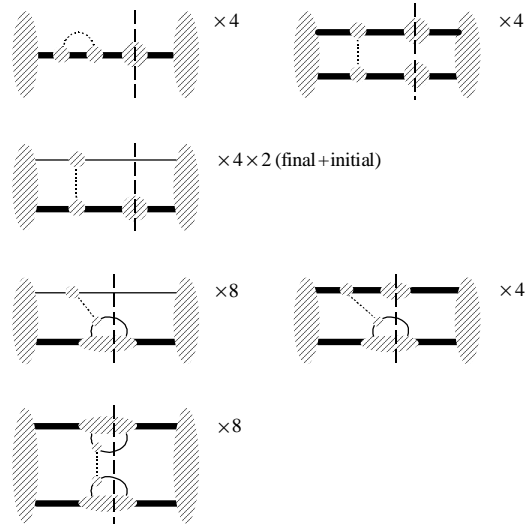

0.7

The thin solid lines correspond to stable "fermions" (the particle's spine is irrelevant because AO commutes with multiplication by polynomials so the numerators may be ignored).

Now I would like to present and discuss an explicit expression for the counterterm for one of the configurations. The configuration is as follows:

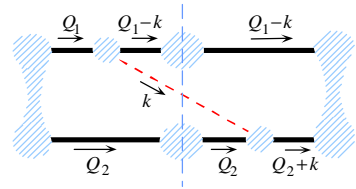

0.8

This corresponds to two chains of unstable propagators with one massless exchange (imagine a $W^+W^-$ pair exchanging a photon; configurations with two massless exchanges may correspond to gluons). For simplicity we consider the case when the masses of the two unstable chains are equal.

The explicit expression for the product of propagators prior to expansion in self-energies is as follows:

$$\Delta(\tau_1;\kappa) W(\tilde{\tau}_{1-};\kappa) \overline{\Delta(\tilde{\tau}_{2+};\kappa)} W(\tau_2;\kappa) \delta_+(k^2) .$$

0.9

The notations are fixed by the correspondence with Fig. 0.8. For instance, $\tau_1 = M^2 - Q_1^2$, etc. Following [1], we denote $\kappa = g^2$, and for the unstable propagator we write $\Delta(\tau;\kappa) = 1/[\tau - \kappa h(\tau) - i\kappa f(\tau)]$, where $h$ and $f$ are the real and imaginary parts of the one-loop self-energy. We will also be using the following notations:

$$h_n = h^{(n)}(\tau)\big|_{\tau=0}, \quad f_n = f^{(n)}(\tau)\big|_{\tau=0}. \qquad 0.10$$

$$\tilde{\gamma} = \gamma + \ln(4\pi) + 2\ln(\kappa f_0/M). \qquad 0.11$$

$$E = \sqrt{(Q_1 + Q_2)^2}/2, \quad Q = \sqrt{E^2 - M^2}. \qquad 0.12$$

The expansion corresponding to the product 0.9 is as follows:

$$(\text{Taylor, VP}) + c_0 \delta(\tau_1)\delta(\tau_2)\delta(k)$$
$$- c_1^{\tau_1} \delta'(\tau_1)\delta(\tau_2)\delta(k) - c_1^{\tau_2} \delta(\tau_1)\delta'(\tau_2)\delta(k)$$
$$- c_1^{\mu} \delta(\tau_1)\delta(\tau_2)\partial_\mu \delta(k) + O(1). \qquad 0.13$$

The first term is the formal Taylor expansion with the products of unstable propagators regulated by VP prescriptions.

The explicit expressions for the coefficients $c$ are as follows. The most cumbersome expression is for the coefficient of the $\delta$-function without derivatives:

$$c_0 = c_0^{\text{NLO}} + c_0^{\text{NNLO}}, \qquad 0.14$$

$$c_0^{\text{NLO}} = \frac{\pi^4}{4\kappa^2 f_0^2 E}\left[\left(\frac{1}{\varepsilon} - \tilde{\gamma}\right)A + B\right], \qquad 0.15$$

$$c_0^{\text{NNLO}} = 2\kappa\left[2h_1 - h_0 f_1 f_0^{-1} + \frac{h_0 + if_0}{4E^2}\right]\cdot c_0^{\text{NLO}}$$
$$+ \frac{\pi^4}{2\kappa f_0^2 E}\left[\frac{h_0 + if_0}{2EM^2}\left(\frac{1}{\varepsilon} - \tilde{\gamma}\right) - \frac{h_1 f_0 + h_0 f_1}{f_0}A\right], \qquad 0.16$$

where

$$A = \frac{1}{Q}\ln\frac{E+Q}{E-Q}, \quad B = \frac{1}{Q}\left[\text{Li}\left(\frac{2Q}{E+Q}\right) - \text{Li}\left(-\frac{2Q}{E-Q}\right)\right]. \qquad 0.17$$

Both $A$ and $B$ are expressed via simple 1-dimensional integrals, and in fact the efforts invested into the obtaining of expressions such as 0.17 are usually wasted because such logarithmic and dilogarithmic contributions tend to cancel in sums over entire families of singular configurations, and such cancellations can be observed already at the stage of 1-dimensional integrals. The reasons behind the cancellations are the same as discussed in [1].

The poles in $\varepsilon$ are a result of the use of dimensional regularization (VP prescriptions alone are insufficient), and the cancellation of (di)logarithms follows the pattern of cancellation of such poles.

For the coefficients of $\delta$-functions with derivatives one has:

$$c_1^{\tau_1} = \kappa(h_0 + if_0)\cdot c_0^{\text{NLO}} + \frac{\pi^4}{\kappa f_0 M^2}\left[\pi - i\left(\frac{1}{\varepsilon} - \tilde{\gamma} + 2\right)\right], \qquad 0.18$$

$$c_1^{\tau_2} = \kappa(h_0 + if_0)\cdot c_0^{\text{NLO}}, \qquad 0.19$$

$$c_1^{\mu} = \left(Q_1^\mu + Q_2^\mu\right)A_+ + \left(Q_1^\mu - Q_2^\mu\right)A_-, \qquad 0.20$$

where

$$A_+ = -\frac{1}{2E^2}\cdot\frac{\pi^5}{2\kappa f_0 M^2}, \qquad 0.21$$

$$A_- = \frac{i}{2Q^2}\left[\kappa f_0 \cdot c_0^{\text{NLO}} - \frac{\pi^4}{2\kappa f_0 M^2}\left(\frac{1}{\varepsilon} - \tilde{\gamma} + 2\right)\right]. \qquad 0.22$$

The conclusions are as follows.

(i) The number of non-trivial counterterms required for writing out explicit expressions for probabilities within the framework of sysPT is fairly large although finite — $O(10)$ rather than $O(100)$ for a typical process with a single massless exchange.

(ii) Calculation of the coefficients of counterterms is quite cumbersome but entirely straightforward and presents no difficulties.

(iii) There occur significant cancellations of logarithmic and dilogarithmic contributions (closely related to cancellations of soft singularities in QED) — and these can be observed prior to the cumbersome explicit evaluation of the integrals.

It may be useful to have a direct all-order verification of the cancellations which ensure gauge invariance without relying on the indirect arguments of [1] but this is not really needed in practical calculations.

*Acknowledgments*. The presented explicit results are based on the calculations performed in collaboration with M. Nekrasov [11]. This work was supported in part by the Russian Foundation for Basic Research (grant 99-02-18365).